\documentclass[preprint]{aastex}
\usepackage{graphicx}
\usepackage{epsf}

\newcommand{\lae}{\lower 2pt \hbox{$\, \buildrel {\scriptstyle <}\over {\scriptstyle\sim}\,$}}
\newcommand{\gae}{\lower 2pt \hbox{$\, \buildrel {\scriptstyle >}\over {\scriptstyle\sim}\,$}}

\def\txt#1{\rm{#1}}

\def\rsun{\ifmmode {\rm R_{\odot}}\else $\rm R_{\odot}$\fi}
\def\msun{\ifmmode {\rm M_{\odot}}\else $\rm M_{\odot}$\fi}
\def\mbh{\ifmmode {\rm M_{\bullet}}\else $\rm M_{\bullet}$\fi}
\def\AU{\ifmmode {{\txt{AU}}}\else AU\fi}
\def\kms{\ifmmode {{\txt{km s^{-1}}}}\else km s$^{-1}$\fi}
\def\yr{\ifmmode {{\txt{yr^{-1}}}}\else yr$^{-1}$\fi}

\def\sstar{S~star}
\def\sstars{S~stars}
\def\abin{a_{\rm bin}}
\def\rclose{r_{\rm close}}

\def\ma{m_1}
\def\mb{m_2}

\def\kdis{k_{\txt{dis}}}

\def\mseed{{M_{\txt{seed}}}}

\def\etidal{{e_{\txt{tidal}}}}

\def\mej{{m_{\txt{ej}}}}

\def\mbintot{m_{\txt{bin}}}
\def\abound{a_{\txt{bnd}}}
\def\ebound{e_{\txt{bnd}}}
\def\abin{a_{\txt{bin}}}
\def\rclose{R_{\txt{close}}}

\def\mavg{m_{\txt{avg}}}

\def\fbin{{f_{\txt{bin}}}}
\def\trel{\tau_{\txt{rel}}}
\def\rtidal{R_{\txt{tidal}}}
\def\period{P}
\def\sgr{Sgr~A$^\ast$}
\def\sigmabh{{\sigma_\bullet}}
\def\rhomax{{\rho_{\txt{max}}}}
\def\mbhunit{{4\!\times\!10^6\,\msun}}

\def\facc{f_{\txt{acc}}}

\begin{document}

\title{Binary Disruption by Massive Black Holes:
Hypervelocity Stars, S Stars, and Tidal Disruption Events}

\author{Benjamin C. Bromley}
\affil{Department of Physics \& Astronomy, University of Utah,
 115 S 1400 E, Rm 201, Salt Lake City, UT 84112}
\email {bromley@physics.utah.edu}

\author{Scott J. Kenyon}
\email{skenyon@cfa.harvard.edu}
\author{Margaret J. Geller}
\email{mgeller@cfa.harvard.edu}
\author{Warren R. Brown}
\affil{Smithsonian Astrophysical Observatory,
60 Garden St., Cambridge, MA 02138}
\email{wbrown@cfa.harvard.edu}

\clearpage

\begin{abstract}
We examine whether disrupted binary stars can fuel black hole growth.
In this mechanism, tidal disruption produces a single hypervelocity 
star (HVS) ejected at high velocity and a former companion star 
bound to the black hole.  After a cluster of bound stars forms, 
orbital diffusion allows the black hole to accrete stars by tidal 
disruption at a rate comparable to the capture rate. In the Milky Way, 
HVSs and the \sstar\ cluster imply similar rates of 
$10^{-5} - 10^{-3}$~\yr\ for binary disruption.  These rates are 
consistent with estimates for the tidal disruption rate in nearby 
galaxies and imply significant black hole growth from disrupted 
binaries on 10~Gyr time scales. 

\end{abstract}

\keywords{
        Galaxy: kinematics and dynamics ---
        Galaxy: halo ---
        Galaxy: stellar content ---
        stars: early-type
}

\section{Introduction}

Massive black holes with masses $\mbh \lesssim 10^9$~\msun\ inhabit the
centers of many galaxies.  It is uncertain how
these objects form and grow. Proposed ideas include the direct collapse 
of a primordial gaseous cloud \citep{eisenhauer05,begelman06}, accretion 
of gas from a surrounding disk \citep{debuhr10}, capture of stellar-mass 
objects from a nuclear cluster \citep{merritt04}, and hierarchical merging 
of black holes \citep{volonteri03}.  For all mechanisms, growing the most 
massive black holes is challenging.  Gaseous accretion flows are limited 
by the need to shed angular momentum for gas to fall efficiently onto the 
black hole.  Pairs of black holes tend to eject material from their vicinity, 
slowing orbit contraction and stalling the growth of either black hole
\citep{miralda00,milo03}.

Recent observations of candidate stellar tidal disruption events 
\citep[TDEs; e.g.,][]{gezari09,vanvelzen11} have renewed interest 
in stellar capture hypotheses.  When a single star wanders inside 
the black hole's tidal radius, the black hole shreds the star.  Some 
material accretes onto the black hole and powers a flare \citep{rees88}.  
Flare evolution depends on properties of the black hole and the shredded 
star \citep[e.g.,][and references therein]{loeb97,strubbe09,lodato11}. 
Observational and theoretical estimates suggest capture rates of 
$10^{-5} - 10^{-3}$ yr$^{-1}$ per nearby galaxy \citep{wang04,vanvelzen11}, 
sufficient to grow a fairly massive black hole on cosmological time scales
\citep[e.g.,][]{merritt04,brockamp11}.

Here, we assess the contribution of binary stars to TDEs and the growth 
of black holes. Binary stars are a significant fraction of all stars 
\citep{abt83,duquennoy91}.  When a binary system wanders too close to 
a black hole, the three-body interaction ejects one binary partner at 
high speeds and leaves the other bound to the black hole \citep{hills88}.  
In addition to predicting the ejected binary components as hypervelocity 
stars (HVSs), the Hills mechanism yields bound stars which serve as a 
mass reservoir for TDEs and black hole growth.  Measurements of both
the bound stars and the HVSs provide direct constraints on the capture
frequency along with robust predictions for the frequency of TDEs 
and the rate of black hole growth in the Galactic Center.

We show that this binary model is consistent with observations of HVSs 
\citep{brown05,brown12b}, and the \sstar\ cluster in the Galactic Center 
\citep{gould03b,oleary08} (\S2).  We use a simple two-body 
relaxation model to show how captured stars can diffuse onto orbits 
that intersect the black hole, leading to a rate for TDEs similar to 
the observations (\S3). Finally, we demonstrate the potential for the 
binary capture mechanism to grow supermassive black holes (\S4).

\section{Binary Disruption Rates from HVSs and S Stars}

To motivate the binary star model, we consider the collision cross-section 
of the black hole, $\sigmabh$. Black holes shred a single star with mass 
$m$ and radius $r$ at the tidal radius,
\begin{eqnarray}\label{eq:rtidal}
\rtidal & \approx & \left(\frac{0.7~\mbh}{m}\right)^{1/3} r 
\approx 0.6-0.7 \left(\frac{\mbh}{4\times 10^6~\msun}\right)^{1/3} 
\left(\frac{r}{\rsun}\right)~{\rm AU}~.
\end{eqnarray}
When $\mbh\le 10^8$~\msun, this interaction occurs outside the
Schwarzschild radius; the black hole accretes a fraction $\facc
\approx 0.25$--$0.50$ of the stellar mass \citep{evans89,ayal00}. 
More massive black holes have Schwarzschild radii exceeding $\rtidal$; 
stars cross the horizon intact and $\facc\equiv 1$.

For encounters with a binary star having semimajor axis $\abin$ and
component masses $\ma$ and $\mb$, the probability of a
capture/ejection is $P_{cap}\approx 1-D/175$ \citep{hills88}, where
$D \approx (\rclose / \abin) f(m, \mbh)$, $\rclose$ is the distance of
closest approach, and $f(m,\mbh)\approx$~1--10 is a function of 
$\mbh$ and $\ma+\mb$.  Conservatively setting $P_{cap}=0.5$,
\begin{equation}
\label{eq:rclose}
\rclose=30-200~\left(\frac{\abin}{\rm 1~AU}\right)~{\rm AU }~.
\end{equation}
Adopting $\sigmabh\approx\pi\rtidal^2\approx\pi
\rclose^2$, interactions with binaries ($\abin\gtrsim 0.1$~AU) are 
$\gtrsim 100$ times more likely than interactions with single stars 
\citep{hopman09,antonini11}.

The \citet{hills88} proposal yields 
(i) ejected stars traveling out into the Galaxy and 
(ii) captured stars bound to the black hole.  For 
$\mbh \sim \mbhunit$ in \sgr\ \citep{eisenhauer05,ghez08}, expected ejection 
speeds are $\sim 1000$--1500~\kms. Deceleration through the Galaxy reduces 
speeds to $\sim 500$--1000~\kms\ at 40--100 kpc \citep{bromley06,kenyon08}.  
Predicted orbits for the resulting bound stars have semimajor axes 
$a\sim 1000$--10000~AU and eccentricities $e\approx$ 0.95--1.0.

Observations reveal both HVSs and captured stars.  \citet{brown05}
discovered SDSS J090745.0$+$024507, a star with a spectral type (B9), 
age ($<350$~Myr), metallicity ([Fe/H]$\sim 0$), radial velocity 
(850~\kms), and distance ($\sim 70$~kpc) consistent with Hills' 
prediction \citep{brown09a}.  Surveys covering roughly 25\% of the sky 
reveal $\sim$ 20 known HVSs with speeds sufficient to escape the Galaxy
\citep{edelmann05,hirsch05,brown09a,brown12b}.
Observed properties of these stars are consistent with a Galactic Center 
origin 
\citep{brown12b}.

At the Galactic Center, the \sstar\ cluster is plausibly composed of
the bound former partners of HVSs
\citep{gould03b,ginsburg06,oleary08}. The $\sim$ 20 massive stars in
this cluster have orbits with $a\lesssim$ 8000~AU and a broad range
of $e$ \citep{eckart97,ghez98,ghez08,genzel10}. The semimajor axes are
consistent with Hills' predictions. Although observed eccentricities
are smaller than predicted, orbital diffusion after capture can reduce
$e$ substantially \citep{perets08c}. The origin of these stars---or
their binary progenitors---is uncertain.  If they formed in young 
stellar disks in the Galactic Center \citep{lockmann08,madigan09}, 
their number count and orbital distribution may represent an atypical 
snapshot of the Galactic Center. A clear connection between the 
\sstars\ and HVSs is key to testing the binary disruption scenario.

Measured production rates for HVSs and \sstars\ suggest a similar
formation mechanism \citep[e.g.,][]{oleary08}.  Known
HVSs have masses of 2.5--3.0~\msun\ and travel times of 60--200~Myr
\citep{brown12b}. The production rate is roughly $1 \times
10^{-6}$~\yr. Correcting this rate for unobserved, lower mass stars
using a standard (differential) initial mass function (IMF) $\xi(m)
\propto m^{-q-1}$ with $q=1.00$--1.35, the production rate for HVSs of
all masses is $2$--$8\times 10^{-5}$~\yr.  \sstars\ have masses
exceeding 5~\msun\ and main sequence lifetimes $t_{ms}\lesssim
100$~Myr \citep{ghez98,genzel10}, yielding a production rate of
$2\times 10^{-7}$~\yr.  Accounting for lower mass stars implies a rate
of 1--$4\times 10^{-5}$~\yr. Within the errors, the rates implied by
HVSs and \sstars\ agree.  The good agreement between the binary disruption
rates derived from two populations with very different lifetimes suggests 
a common, steady production mechanism. 

For a more realistic estimate of the binary disruption rate from HVSs
and \sstars, we assume continuous star formation \citep{figer04}, and 
steady, random scattering of binary stars toward the Galactic Center. 
Thus the pool of progenitors, accumulated over 10~Gyr, contains more 
older, lower mass stars than predicted by the IMF. To compensate for
this population, we adopt $t_{ms}\lesssim$~500~Myr for HVSs and
$t_{ms}\lesssim$~100~Myr for \sstars. The inferred production rates 
increase by $\sim 10$~Gyr/500~Myr~=~20 for HVSs and 
$\sim 10$~Gyr/100~Myr~=~100 for \sstars. Thus, the maximum rate for
binary disruptions in the Galactic Center is roughly $1$--$2\times
10^{-3}$~\yr.

This range of rates agrees 
with predictions. In simple models, binaries interact with the 
central black hole at a rate of $n\left<v f_{\txt{g}}\sigmabh\right>$, where 
$n$ is the number density of binaries, $v\sim 100$~\kms\ is a characteristic 
speed \citep{figer03}, and $f_{\txt{g}}\sigmabh\sim 1\times 10^8\,\AU^2$
is the cross-sectional area for $\rclose \sim$ 100~AU
(eq. [\ref{eq:rclose}]) and a gravitational focusing factor
$f_{\txt{g}}$.  For a central mass density $\rho_0$, 
$n\approx\fbin\rho_0/\mavg\,(1+\fbin)$, where $\fbin$ is the fraction of 
stars in short period binaries and $\mavg \approx$ 0.3 \msun\ is the average 
mass of a star.  In the solar neighborhood, the fraction of stars in binaries
of all periods is $\fbin=0.5$--0.7 \citep{abt83}.  For the short period 
binaries likely to produce HVSs ($P\lesssim$~1~yr), $\fbin\approx$~0.1 
\citep{yu03}.  The total disruption rate is then
\citep[see also][]{hills88,yu03}
\begin{eqnarray}\label{eq:kdis}
\kdis &\approx& \frac{\fbin}{1+\fbin}\frac{\rho_0}{\mavg}
\left< v f_{\txt{g}}\sigmabh\right> \\
\ \ &\approx&  10^{-3}~\mbox{yr}^{-1}
\end{eqnarray}
for $\rho_0=1.4\times 10^4$~\msun\ pc$^{-3}$ \citep[e.g.,][]{bromley06}
and $\fbin=0.1$. This rate agrees with the maximum rate of binary disruption 
implied by observations.

The maximum disruption rate assumes that scattering processes rapidly
fill the ``loss cone,'' the subset of Galactic Center orbits that pass
close to the black hole.  If scattering supplies binaries to the loss
cone more slowly than disruption removes them, the disruption rate is
comparable to the scattering rate. When two-body relaxation fills the 
loss cone, \citet{yu03} derive $\kdis\approx 10^{-5}$~\yr\ for
$\fbin=0.1$. Because other processes can fill the loss cone, this
estimate is probably a lower limit
\citep{merritt04,hopman06,perets07}.  When the gravitational potential 
is nonaxisymmetric and orbits are chaotic, \citet{merritt04} derive 
factor of ten larger scattering rates. The expected disruption rate is 
then $\kdis\approx 10^{-4}$~\yr. Taken together, the broad range of 
theoretical estimates, $\kdis\sim 10^{-5}$--$10^{-3}$~\yr, is consistent 
with the binary disruption rates implied by observations.

\section{Long-term Orbital Evolution of the S~Stars and Tidal Disruption Events}

The inferred binary disruption rate agrees with estimates for TDEs
from nearby galaxies. Observations of two candidate TDEs from
long-term observations of galaxies in the Sloan Digital Sky Survey
suggest one TDE per galaxy every $\sim 10^5$ yr \citep{vanvelzen11}.
Candidates from GALEX data indicate a similar frequency
\citep{gezari09}.  The good agreement between the rates of binary
disruption and stellar disruption plausibly suggests a common origin. 
To explore this connection, we consider the evolution of the orbits 
of bound stars.

Over 10~Gyr, binary disruption places $\sim 10^5$--$10^7$ bound stars in orbit 
around the central black hole. Just after capture, each star has a semimajor 
axis and eccentricity
\begin{eqnarray}
\abound &\sim& 1100 \ 
\left(\frac{\abin}{0.1~\AU}\right)\left(\frac{1~\msun}{\mej}\right)\left(\frac{\mbintot}{2~\msun}\right)^{1/3}~\AU,
\\
\ebound &=& 1 - 0.005 \left(\frac{\rclose}{10~\AU}\right)\left(\frac{1100~\AU}{\abound}\right)~,
\end{eqnarray}
for $\mbh=4\times 10^6$ \msun.  These equations follow from
conservation of energy and angular momentum, with coefficients from
numerical simulations of binary disruptions
\cite[e.g.,][]{hills88,bromley06}.

Gravitational interactions among the bound stars produce changes in $\abound$ 
and $\ebound$ on a relaxation time $\trel$. Stars with $t_{ms} \lesssim \trel$ 
explode as supernovae or evolve into red giants and white dwarfs
\citep{frank76,hils95,hopman06,merritt10}. Less massive stars may evolve onto 
orbits with periastron distances $a_{peri} \lesssim$ $\rtidal$.  These stars 
produce TDEs and accrete onto the black hole \citep{rees88}.

The rate of TDEs depends on $\trel$. If two-body
scattering dominates, the relaxation time is
\begin{eqnarray}\label{eq:trel}
\trel &\approx&\frac{0.33 M^2}{\mavg^2 N\ln(0.4 N)}\frac{P}{2\pi} 
\\
\ &\approx&130\left(\frac{\mbh}{4\!\times\!10^6\,\msun}\right)^{\!\!5/6}\!\!
\left[\frac{N\!\cdot\!\ln(0.4 N)}{10^6\!\cdot\!12.9}\right]^{\!\!-1}~\rm{Myr}  
\end{eqnarray}
when $\sim 10^6$ stars with an average mass of 0.3~\msun\ orbit at
$\sim 5000$~\AU\ with orbital period $\period$.  Other processes, 
such as resonant relaxation, probably shorten the relaxation time 
\citep[e.g.,][]{hopman06}.  Captured stars wander through 
$0\leq e\leq 1$ on this time scale.  

The relaxation time sets two properties for the bound population. 
Stars with $t_{ms}\lesssim\trel$ ($m\gtrsim 2$--$4 \msun$) evolve 
into supernovae or white dwarfs before they encounter the black hole. 
For a normal IMF, the fraction of stars that suffer this fate is a 
few per cent.  The rest of the population establishes an approximate 
steady-state, where the black hole accretes a star for every captured 
star. For the current mass of \sgr, this process requires $\sim$ 
0.1--1~Gyr. After this period, there is a tidal disruption event 
every $10^3$--$10^5$ yr.

Numerical simulations do not consider the long-term evolution of 
$\sim 10^5$--$10^7$ low mass stars bound to a central black hole. 
However, smaller simulations confirm the general features of this
evolution \citep[e.g.,][]{perets08c}.  Gravitational interactions
lead to global changes in $a$ and $e$ on the relaxation time $\trel$. 
At least 20\% to 30\% of bound stars suffer TDEs.  Thus, \sgr\ should 
accrete a 0.1--0.5 \msun\ star every $10^3$--$10^5$~yr.

\section{Growth of the Central Black Hole}

The upper limit on the tidal disruption rate can add mass comparable 
to the total mass of \sgr.  To explore how binary disruption impacts 
$\mbh$, we consider a model where the capture rate is the collision 
rate in equation (\ref{eq:kdis}).  For a constant central 
density $\rho_0=10^4$~\msun~pc$^{-3}$, velocity $v=100$~\kms, and 
binary fraction $\fbin=0.1$, our time-dependent calculations begin 
with an initial $\mbh=\mseed$.  Captured stars accumulate until time 
$t=\trel$ and then accrete with efficiency $\facc$ onto the black hole.  
The growth of $\mbh$ depends only on the initial mass and the accretion 
efficiency.

\begin{figure}
\centerline{\includegraphics[width=5.0in]{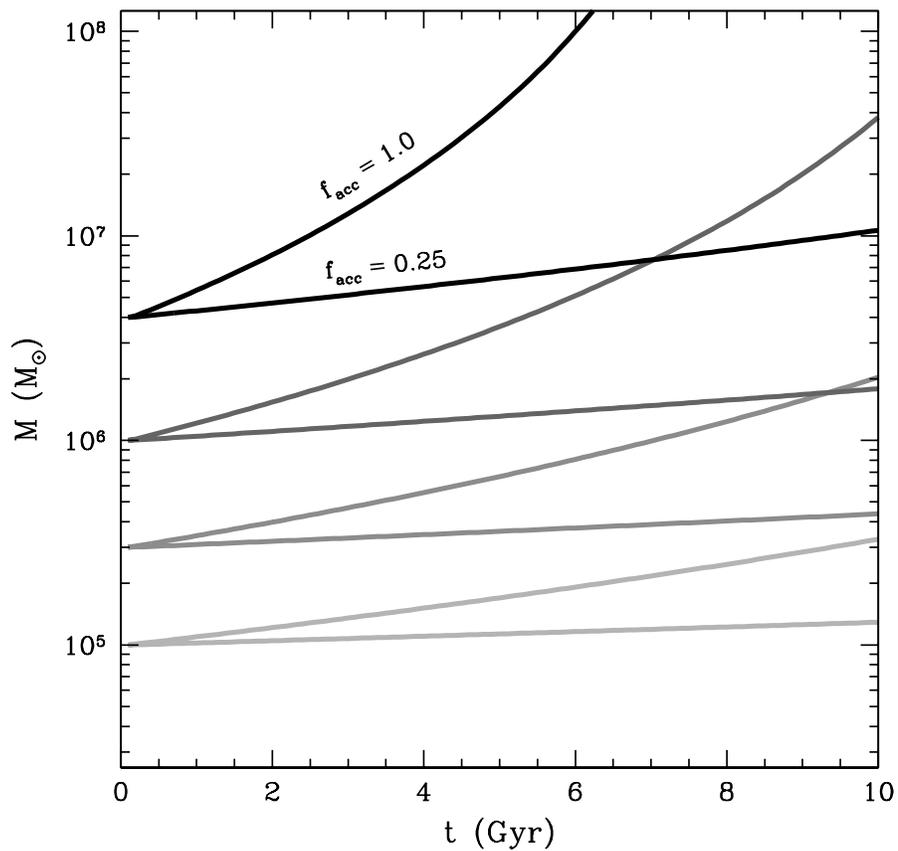}}
\caption{Growth of a black hole from a seed mass. Curves
correspond to seed masses between $10^5$~\msun\ and 
$3.5\times 10^6$~\msun. The lower (upper) curves have 
$\facc=0.25$ ($\facc=1$).  
\label{fig:bhgrow}
}
\end{figure}

Figure~\ref{fig:bhgrow} shows results for several seed masses and 
$\facc=0.25$ and 1.00.  In Figure~\ref{fig:bhgrow}, small seed masses 
and $\facc\le$ 1 yield modest growth rates over 10~Gyr.  Larger seed 
masses of a few $\times 10^5$~\msun\ yield $\mbh\approx\mbhunit$ 
within 10~Gyr.

To show how black hole growth depends on $\rho_0$ and $\fbin$, we define 
a growth parameter
\begin{equation}
\label{eq:eta}
\eta = \left({\facc\over 1}\right)\left({\rho_c\over\rho_0}\right)\left({\fbin \over 0.1}\right)
\end{equation}
where $\rho_c$ is the central mass density near the black hole. Thus,
$\eta=1$ corresponds to models with the Milky Way's central stellar 
mass density. Calculations with $\eta>1$ ($\eta<1$) yield larger 
(smaller) growth rates.

\begin{figure}
\centerline{\includegraphics[width=5.0in]{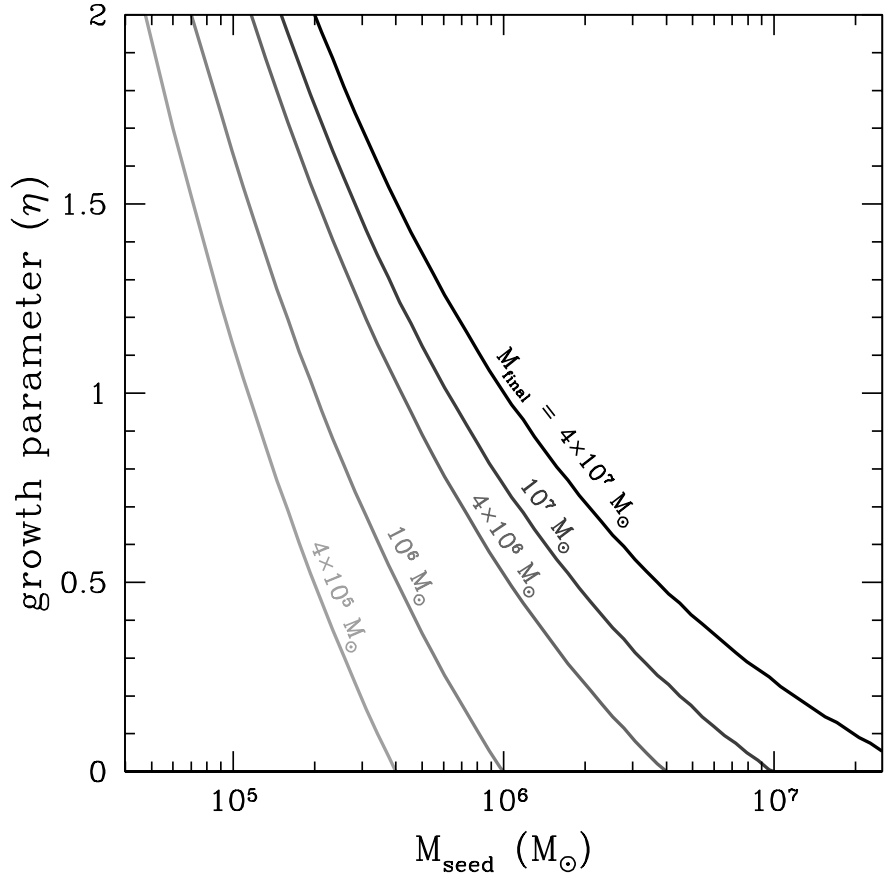}}
\caption{Tracks of constant final $\mbh$ as a function of 
the growth parameter $\eta$ and the initial $\mbh$. 
For the accretion rates expected in the Galactic Center, 
initial masses of $\sim 10^6$ \msun\ yield the mass of 
\sgr\ in 10~Gyr.
\label{fig:Mfacc}
}
\end{figure}

Figure~\ref{fig:Mfacc} displays $\mbh$ at 10 Gyr as a function of $\eta$ 
and the initial seed mass.  In this model, massive 
black holes require large accretion efficiency, large central density, 
and a large fraction of short period binaries.  Low values for $\eta$ 
yield low $\mbh$.  Figs.~\ref{fig:bhgrow}--\ref{fig:Mfacc} imply that 
the current binary disruption rate implies a substantially lower mass 
for \sgr\ in the past.

To explore black hole growth in more detail, we add the evolution of 
$\rtidal$ (eq.~[\ref{eq:rtidal}]) and $\rclose$ (eq.~[\ref{eq:rclose}]) 
as the black hole grows in mass.  Large gravitational focusing factors 
imply that the disruption rate is independent of $v$ (eq.~[\ref{eq:kdis}]).  
We also allow the central density to vary in time, adopting 
$\rho(t)\propto\mbh^{-\gamma}$ and $\gamma=1$--2,
as observed in elliptical galaxies \citep{faber97}. This density is
bounded by a maximum initial density of $\rhomax=10^6$~\msun~pc$^{-3}$.

In this model, the mass accretion rate is 
\begin{equation}\label{fig:dmbhdt}
\frac{d\mbh}{dt}=\frac{2\facc\mavg\,(1+\etidal)}{\trel},
\end{equation}
which includes the rate of diffusion into orbits that exceed
$\etidal$, the eccentricity corresponding to a perihelion distance of
$\rtidal$.  The resulting growth of the black hole is not sensitive to
the exact form of this equation. As in the simple model
(Figs.~\ref{fig:bhgrow}--\ref{fig:Mfacc}), the rate invariably adjusts
to achieve a steady state with a balance between capture of stars and
the tidal shredding by the black hole.

Although there may be no direct causal connection between the black
hole and the central density outside of its sphere of influence, our
approach implies that the central density diminishes in time as $\mbh$
grows.  Once the central density falls below $\rhomax$, the numerical 
results suggest that the central density declines with time roughly as 
$\rho(t)\sim t^{-\beta}$, where $\beta\sim 1$.

\begin{figure}
\centerline{\includegraphics[width=5.0in]{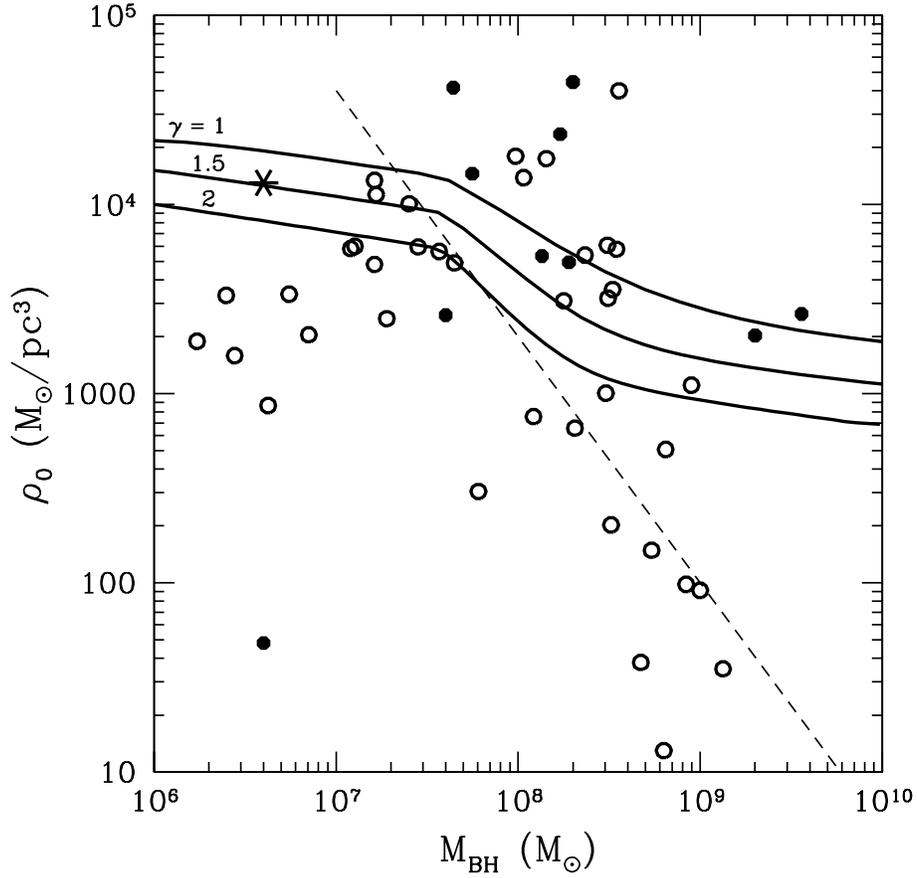}}
\caption{ Central stellar mass density as a function of final black
  hole mass.  For models with a seed mass of $10^5$~\msun,
  $\facc=0.25$, and $\rho(t)\propto\mbh^{-\gamma}$, the upper
  ($\gamma=1$), middle ($\gamma=1.5$), and lower ($\gamma=2$) solid
  lines show predicted relations between $\rho_0$ and $\mbh$ at 10
  Gyr.  We generate each curve by varying the {\em initial} density
  $\rho(t=0)$, and letting $\rho(t)$ evolve as $\mbh$ grows. Thus,
  while a large initial density yields a large black hole mass, the 
  present-day density $\rho_0$ is driven to a small value if
  $\gamma>0$.  Observations support this correlation.  Filled circles 
  show data from the \citet{faber97} sample with $\mbh$ from
  \citet{aller07}. The open circles are from the same sample, but with
  the assumption that $\mbh\sim v^4$, where $v$ is the measured
  velocity dispersion \citep{tremaine02}.  The dashed line is a
  power-law relation with $\gamma=1.3$, representative of galaxies in
  the local universe.
\label{fig:rhombh}
}
\end{figure}

Figure \ref{fig:rhombh} summarizes our results.  For an initial black
hole seed mass of $10^5$~\msun, the two curves show the central
density as a function of black hole mass at $t=10$~Gyr. For any
$\mbh$ at 10~Gyr, models with $\gamma=1$
require larger central stellar mass densities than models with
$\gamma=2$.  Observations show considerable scatter about the
general trend derived from local galaxies \citep{faber97}. Although we
cannot discriminate among plausible sets of variables, the growth
models yield a fair representation of the trend among galaxies with
large central stellar mass densities. These models also appear to fare
better than the power-law correlation $\rho\propto\mbh^{-1.3}$ from
dynamical arguments \citep[dashed line in the figure;][]{faber97}.

\section{Summary}

We develop a model for binary disruption in the Galactic Center.  The
model relates observations of HVSs in the halo and the \sstar\ cluster
in the Galactic Center, TDEs in nearby galaxies,
and the growth of central black holes.  The black hole in the Galactic 
Center disrupts close binaries at distances $\lesssim\rclose$ 
(eq.~[\ref{eq:rclose}]). One companion (the HVS) is ejected at high 
speeds; the other (\sstar) becomes bound to a black hole.  Among the 
bound stars, dynamical interactions push some stars inside the tidal 
radius (eq.~[\ref{eq:rtidal}]). Shredding of stars inside $\rtidal$
produces TDEs. After a relaxation time
(eq.~[\ref{eq:trel}]), the rate of tidal disruptions roughly equals 
the capture rate.  For large enough capture rates, binary disruptions 
add significant mass to the central black hole.

Current data suggest this picture is plausible.  Observations of HVSs
and \sstars\ yield consistent production rates of
$10^{-5}$--$10^{-3}$~\yr; these rates agree with theoretical
estimates. For $\trel\approx$ 0.1--1~Gyr, the expected rate for TDEs,
$10^{-5}$--$10^{-3}$~\yr, is close to current estimates from small
samples.  More plentiful single stars may contribute to TDEs but the
rate is at or below the level of the binary disruption channel
\citep{brockamp11}.  For rates associated with binary disruption,
\sgr\ has grown by at least a factor of 2--4 in the past 5--10 Gyr. At
higher rates, binary disruption can produce black holes with
$\mbh\gtrsim 10^8$~\msun. This growth model yields clear relations
between the final $\mbh$ and the central stellar mass density. These
relations are reasonably consistent with existing data.

This model is attractive in connecting HVS and \sstars\ in the Milky Way 
to TDEs in nearby galaxies.  To test this connection,
$N$-body simulations with larger numbers of stars bound to the black hole 
would improve estimates of the relaxation time and relations between
the capture rate and the disruption rate.  Theoretical investigations
of TDEs involving the bound former partners of disrupted binaries might 
yield different light curves which could be
tested by observations.

Several observations also provide essential tests.

\begin{itemize}

\item Extending HVS surveys to the southern sky with {\it SkyMapper}
  would improve estimates of the HVS production rate. In the future,
  {\it LSST} should detect candidates with lower mass and at larger
  distances in the halo. Spectroscopic confirmation with current
  6--10-m telescopes and proposed 25--50-m telescopes would yield even
  better constraints on the HVS population.

\item Discovery of lower mass stars among the bound star population
  yields another test of the binary disruption hypothesis. Deeper
  observations of the Galactic Center with existing large telescopes,
  the {\it James Webb Space Telescope}, and planned 25--50-m
  telescopes should reveal this population.

\item Larger samples of candidate TDEs enable
  better comparisons of the binary disruption rate in the Milky Way
  with tidal disruption rates in external galaxies. Aside from
  existing surveys with the SDSS and {\it Pan-Starrs}, surveys with
  {\it SkyMapper} and {\it LSST} will yield strong constraints. In any
  picture, low mass stars should produce most TDEs.  With constraints 
  on the black hole mass, deriving a
  luminosity function for this process might place estimates on the
  mass function of disrupted stars. Rates as a function of central
  stellar mass density tests theories for the capture rate.

\end{itemize}

Calculations of black hole growth by tidal disruption predict a
correlation between the stellar mass (velocity dispersion) and the
final $\mbh$.  Although our models cannot predict the 
observed $\mbh\propto\sigma^4$ relation, more sophisticated 
calculations could reveal $\mbh$-$\sigma$ correlations which might 
be tested observationally.


\begin{thebibliography}{50}
\expandafter\ifx\csname natexlab\endcsname\relax\def\natexlab#1{#1}\fi

\bibitem[{{Abt}(1983)}]{abt83}
{Abt}, H.~A. 1983, \araa, 21, 343

\bibitem[{{Aller} \& {Richstone}(2007)}]{aller07}
{Aller}, M.~C., \& {Richstone}, D.~O. 2007, \apj, 665, 120

\bibitem[{{Antonini} {et~al.}(2011){Antonini}, {Lombardi}, \&
  {Merritt}}]{antonini11}
{Antonini}, F., {Lombardi}, Jr., J.~C., \& {Merritt}, D. 2011, \apj, 731, 128

\bibitem[{{Ayal} {et~al.}(2000){Ayal}, {Livio}, \& {Piran}}]{ayal00}
{Ayal}, S., {Livio}, M., \& {Piran}, T. 2000, \apj, 545, 772

\bibitem[{{Begelman} {et~al.}(2006){Begelman}, {Volonteri}, \&
  {Rees}}]{begelman06}
{Begelman}, M.~C., {Volonteri}, M., \& {Rees}, M.~J. 2006, \mnras, 370, 289

\bibitem[{{Brockamp} {et~al.}(2011)}]{brockamp11} {Brockamp}, M., 
{Baumgardt}, H., \& {Kroupa}, P.\ 2011, \mnras, 418, 1308 

\bibitem[{{Bromley} {et~al.}(2006){Bromley}, {Kenyon}, {Geller}, {Barcikowski},
  {Brown}, \& {Kurtz}}]{bromley06}
{Bromley}, B.~C., {Kenyon}, S.~J., {Geller}, M.~J., {Barcikowski}, E., {Brown},
  W.~R., \& {Kurtz}, M.~J. 2006, \apj, 653, 1194

\bibitem[{{Brown} {et~al.}(2009){Brown}, {Geller}, \& {Kenyon}}]{brown09a}
{Brown}, W.~R., {Geller}, M.~J., \& {Kenyon}, S.~J. 2009, \apj, 690, 1639

\bibitem[{{Brown} {et~al.}(2012){Brown}, {Geller}, \& {Kenyon}}]{brown12b}
---. 2012, arXiv:1203.3543 (accepted in \apj) 

\bibitem[{{Brown} {et~al.}(2005){Brown}, {Geller}, {Kenyon}, \&
  {Kurtz}}]{brown05}
{Brown}, W.~R., {Geller}, M.~J., {Kenyon}, S.~J., \& {Kurtz}, M.~J. 2005,
  \apjl, 622, L33

\bibitem[{{Debuhr} {et~al.}(2010){Debuhr}, {Quataert}, {Ma}, \&
  {Hopkins}}]{debuhr10}
{Debuhr}, J., {Quataert}, E., {Ma}, C.-P., \& {Hopkins}, P. 2010, \mnras, 406,
  L55

\bibitem[{{Duquennoy} \& {Mayor}(1991)}]{duquennoy91}
{Duquennoy}, A., \& {Mayor}, M. 1991, \aap, 248, 485

\bibitem[{{Eckart} \& {Genzel}(1997)}]{eckart97}
{Eckart}, A., \& {Genzel}, R. 1997, \mnras, 284, 576

\bibitem[{{Edelmann} {et~al.}(2005){Edelmann}, {Napiwotzki}, {Heber},
  {Christlieb}, \& {Reimers}}]{edelmann05}
{Edelmann}, H., {Napiwotzki}, R., {Heber}, U., {Christlieb}, N., \& {Reimers},
  D. 2005, \apjl, 634, L181

\bibitem[{{Eisenhauer} {et~al.}(2005)}]{eisenhauer05}
{Eisenhauer}, F., {et~al.} 2005, \apj, 628, 246

\bibitem[{{Evans} \& {Kochanek}(1989)}]{evans89}
{Evans}, C.~R., \& {Kochanek}, C.~S. 1989, \apjl, 346, L13

\bibitem[{{Faber} {et~al.}(1997){Faber}, {Tremaine}, {Ajhar}, {Byun},
  {Dressler}, {Gebhardt}, {Grillmair}, {Kormendy}, {Lauer}, \&
  {Richstone}}]{faber97}
{Faber}, S.~M., {et~al.} 1997, \aj, 114, 1771

\bibitem[{{Figer} {et~al.}(2003){Figer}, {Gilmore}, {Kim}, {Morris}, {Becklin},
  {McLean}, {Gilbert}, {Graham}, {Larkin}, {Levenson}, \& {Teplitz}}]{figer03}
{Figer}, D.~F., {et~al.} 2003, \apj, 599, 1139

\bibitem[{{Figer} {et~al.}(2004)}]{figer04} Figer, D.~F., Rich, 
R.~M., Kim, S.~S., Morris, M., \& Serabyn, E.\ 2004, \apj, 601, 319 

\bibitem[{{Frank} \& {Rees}(1976)}]{frank76}
{Frank}, J., \& {Rees}, M.~J. 1976, \mnras, 176, 633

\bibitem[{{Genzel} {et~al.}(2010){Genzel}, {Eisenhauer}, \&
  {Gillessen}}]{genzel10}
{Genzel}, R., {Eisenhauer}, F., \& {Gillessen}, S. 2010, Reviews of Modern
  Physics, 82, 3121

\bibitem[{{Gezari} {et~al.}(2009){Gezari}, {Heckman}, {Cenko}, {Eracleous},
  {Forster}, {Gon{\c c}alves}, {Martin}, {Morrissey}, {Neff}, {Seibert},
  {Schiminovich}, \& {Wyder}}]{gezari09}
{Gezari}, S., {et~al.} 2009, \apj, 698, 1367

\bibitem[{{Ghez} {et~al.}(1998){Ghez}, {Klein}, {Morris}, \&
  {Becklin}}]{ghez98}
{Ghez}, A.~M., {Klein}, B.~L., {Morris}, M., \& {Becklin}, E.~E. 1998, \apj,
  509, 678

\bibitem[{{Ghez} {et~al.}(2008){Ghez}, {Salim}, {Weinberg}, {et~al.}}]{ghez08}
{Ghez}, A.~M., {Salim}, S., {Weinberg}, N.~N., {et~al.} 2008, \apj, 689, 1044

\bibitem[{{Ginsburg} \& {Loeb}(2006)}]{ginsburg06}
{Ginsburg}, I., \& {Loeb}, A. 2006, \mnras, 368, 221

\bibitem[{{Gould} \& {Quillen}(2003)}]{gould03b}
{Gould}, A., \& {Quillen}, A.~C. 2003, \apj, 592, 935

\bibitem[{{Hills}(1988)}]{hills88}
{Hills}, J.~G. 1988, \nat, 331, 687

\bibitem[{{Hils} \& {Bender}(1995)}]{hils95}
{Hils}, D., \& {Bender}, P.~L. 1995, \apjl, 445, L7

\bibitem[{{Hirsch} {et~al.}(2005){Hirsch}, {Heber}, {O'Toole}, \&
  {Bresolin}}]{hirsch05}
{Hirsch}, H.~A., {Heber}, U., {O'Toole}, S.~J., \& {Bresolin}, F. 2005, \aap,
  444, L61

\bibitem[{{Hopman}(2009)}]{hopman09}
{Hopman}, C. 2009, \apj, 700, 1933

\bibitem[{{Hopman} \& {Alexander}(2006)}]{hopman06}
{Hopman}, C., \& {Alexander}, T. 2006, \apj, 645, 1152

\bibitem[{{Kenyon} {et~al.}(2008){Kenyon}, {Bromley}, {Geller}, \&
  {Brown}}]{kenyon08}
{Kenyon}, S.~J., {Bromley}, B.~C., {Geller}, M.~J., \& {Brown}, W.~R. 2008,
  \apj, 680, 312


\bibitem[L{\"o}ckmann et al.(2008)]{lockmann08} L{\"o}ckmann, U., 
Baumgardt, H., \& Kroupa, P.\ 2008, \apjl, 683, L151 


\bibitem[{{Lodato} \& {Rossi}(2011)}]{lodato11}
{Lodato}, G., \& {Rossi}, E.~M. 2011, \mnras, 410, 359

\bibitem[{{Loeb} \& {Ulmer}(1997)}]{loeb97}
{Loeb}, A., \& {Ulmer}, A. 1997, \apj, 489, 573

\bibitem[Madigan et al.(2009)]{madigan09} Madigan, A.-M., Levin, 
Y., \& Hopman, C.\ 2009, \apjl, 697, L44 

\bibitem[{{Merritt}(2010)}]{merritt10}
{Merritt}, D. 2010, \apj, 718, 739

\bibitem[{{Merritt} \& {Poon}(2004)}]{merritt04}
{Merritt}, D., \& {Poon}, M.~Y. 2004, \apj, 606, 788

\bibitem[{{Milosavljevi{\'c}} \& {Merritt}(2003)}]{milo03}
{Milosavljevi{\'c}}, M., \& {Merritt}, D. 2003, in American Institute of
  Physics Conference Series, Vol. 686, The Astrophysics of Gravitational Wave
  Sources, ed. {J.~M.~Centrella}, 201--210

\bibitem[{{Miralda-Escud{\'e}} \& {Gould}(2000)}]{miralda00}
{Miralda-Escud{\'e}}, J., \& {Gould}, A. 2000, \apj, 545, 847

\bibitem[{{O'Leary} \& {Loeb}(2008)}]{oleary08}
{O'Leary}, R.~M., \& {Loeb}, A. 2008, \mnras, 383, 86


\bibitem[{{Perets} {et~al.}(2009){Perets}, {Gualandris}, {Merritt}, \&
  {Alexander}}]{perets08c}
{Perets}, H.~B., {Gualandris}, A., {Merritt}, D., \& {Alexander}, T. 2009,
  \apj, 702, 884

\bibitem[{{Perets} {et~al.}(2007){Perets}, {Hopman}, \& {Alexander}}]{perets07}
{Perets}, H.~B., {Hopman}, C., \& {Alexander}, T. 2007, \apj, 656, 709

\bibitem[{{Rees}(1988)}]{rees88}
{Rees}, M.~J. 1988, \nat, 333, 523

\bibitem[{{Strubbe} \& {Quataert}(2009)}]{strubbe09}
{Strubbe}, L.~E., \& {Quataert}, E. 2009, \mnras, 400, 2070

\bibitem[{{Tremaine} {et~al.}(2002){Tremaine}, {Gebhardt}, {Bender}, {Bower},
  {Dressler}, {Faber}, {Filippenko}, {Green}, {Grillmair}, {Ho}, {Kormendy},
  {Lauer}, {Magorrian}, {Pinkney}, \& {Richstone}}]{tremaine02}
{Tremaine}, S., {et~al.} 2002, \apj, 574, 740

\bibitem[{{van Velzen} {et~al.}(2011){van Velzen}, {Farrar}, {Gezari},
  {Morrell}, {Zaritsky}, {{\"O}stman}, {Smith}, {Gelfand}, \&
  {Drake}}]{vanvelzen11}
{van Velzen}, S., {et~al.} 2011, \apj, 741, 73

\bibitem[{{Volonteri} {et~al.}(2003){Volonteri}, {Haardt}, \&
  {Madau}}]{volonteri03}
{Volonteri}, M., {Haardt}, F., \& {Madau}, P. 2003, \apj, 582, 559

\bibitem[{{Wang} \& {Merritt}(2004)}]{wang04}
{Wang}, J., \& {Merritt}, D. 2004, \apj, 600, 149

\bibitem[{{Yu} \& {Tremaine}(2003)}]{yu03}
{Yu}, Q., \& {Tremaine}, S. 2003, \apj, 599, 1129

\end{thebibliography}

\clearpage

\end{document}